\documentclass[prl,twocolumn,showpacs,floatfix,amsfonts]{revtex4}
\usepackage{graphicx,graphics,color,epsfig}
\usepackage{bm}
\usepackage{amsmath}
\usepackage{amssymb}

\begin{document}
\preprint{}

\title{LDOS modulations in cuprate superconductors with competing AF
order: the temperature effect}
\author{Hong-Yi Chen and C.S. Ting}
\affiliation{Texas Center for Superconductivity and Advanced Material,
and Department of Physics, University of Houston, Houston, TX 77204}

\begin{abstract}
Based upon a phenomenological $t-t'-U-V$ model and using
Bogoliubov-de Gennes equations, we found that near the optimal doping
$\delta=0.15$ at low temperature ($T$), only the pure d-wave
superconductivity (dSC) prevails and the antiferromagnetic (AF)
order is completely suppressed. However, at higher $T$ the AF order
with stripe modulation and the accompanying charge density wave
(CDW) emerge, and they could exist even above the
superconducting transition temperature. This implies that the
existence of the CDW depends critically on the presence of the AF
order, not so much on the dSC. The LDOS (local density of states)
image indicates that the stripe modulation has an energy
independent spacing of $5a/4a$ spreading over a $24a \times 48a$
lattice, corresponding to an average periodicity $4.8a$. This
result may be relevant to the recent STM experiment [Vershinin $et\;al.$, 
Science {\bf 303}, 1995 (2004)].

\end{abstract}


\maketitle

Recently, STM measurement by Vershinin $et\;al.$
\cite{vershinin303} on slightly underdoped BSCCO indicated that
the electronic states at low energies and at a temperature $T$
higher than the superconducting transition temperature ($T_c$) in
the pseudogap region exhibit an energy-independent spatial
modulation which resulting to a checkerboard pattern with
incommensurate periodicity $4.7a \pm 0.2$ ($a$ is the lattice
constant). At very low temperature, however, no such pattern
has been detected \cite{vershinin303}, in agreement with previous
measurements \cite{hoffman297,mcelroy422}. So far the reasons why
this STM pattern at low temperature
looks so different from that at high temperature, and why the
noninteger incommensurate periodicity could occurred in a underline 
CuO$_2$ lattice are outstanding questions which have not been 
addressed in the existing literatures. In the present paper we are 
trying to understand these issues by adopting the idea of the $d$-wave
superconductivity (dSC) with a competing antiferromagnetic (AF)
order for cuprate superconductors
\cite{balents12,sachdev286,himeda60,martin14,daul84,kivelson75}, 
and to examine the formation of the AF order and the accompanying
charge density wave (CDW) at finite temperature. Employing the
phenomenological $t-t'-U-V$ model with proper chosen parameters, we show 
that at low temperature only dSC prevails in our system and the AF order 
is completely suppressed. The LDOS (local density of states) image is 
featureless. 
At higher temperature, it is found that the AF order with stripe 
modulation, which is referred also as the spin density wave (SDW), and the 
accompanying charge density wave (CDW)  may show up and they could even 
persist at temperatures above the BCS superconducting transition 
temperature $T_c^{BCS}$. In the presence of SDW, we show the LDOS image to 
have an energy-independent stripe modulation with spacing of $5a/4a$ 
spreading over a $24a \times 48a$ lattice. According to the Fourier 
analysis of the LDOS image, an average periodicity $4.8a$ could be 
assigned for 
the stripe modulation. If the doubly degenerate states of $x$-and
$y$-oriented stripes have the same probability to appear
in the time interval when the experiments is performed, the
combined LDOS image should have a checkerboard pattern of $4.8a \times
4.8a$ structure. This result is in agreement with  the experiment
by Vershinin $et\;al.$ \cite{vershinin303}. In addition, the temperature 
dependent behavior of the normalized LDOS, exhibits the "pseudogap"-like 
characteristics \cite{kugler86} originated in the emergence of 
the SDW at higher $T$.

To model these observed phenomena, we employ an effective
mean-field $t-t'-U-V$ Hamiltonian by assuming that the on-site
repulsion $U$ is responsible for the competing antiferromagnetism
and the nearest-neighbor attraction $V$ causes the $d$-wave
superconducting pairing

\begin{eqnarray}
{\bf H}&=&-\sum_{{\bf ij}\sigma} t_{\bf ij} c_{{\bf
i}\sigma}^{\dagger}c_{{\bf j}\sigma}
+\sum_{{\bf i}\sigma} ( U\langle n_{{\bf i}\bar{\sigma}}\rangle - \mu )
c_{{\bf i}\sigma}^{\dagger}c_{{\bf i}\sigma} \nonumber \\
&&+\sum_{\bf ij} (\Delta_{\bf ij} c_{{\bf i}\uparrow}^{\dagger}
c_{{\bf j}\downarrow}^{\dagger} +\Delta_{\bf ij}^{*} c_{{\bf
j}\downarrow} c_{{\bf i}\uparrow} )\;,
\end{eqnarray}
where $t_{\bf ij}$ is the hopping integral, $\mu$ is the chemical
potential, and $\Delta_{\bf ij}=\frac{V}{2}\langle c_{{\bf i}\uparrow}
c_{{\bf j}\downarrow}-c_{{\bf i}\downarrow}c_{{\bf j}\uparrow}\rangle$ is
the spin-singlet $d$-wave bond order parameter. The Hamiltonian above
shall be diagonalized by using Bogoliubov-de Gennes' (BdG) equations,
\begin{eqnarray}
\sum_{\bf j}^N \left(\begin{array}{cc}
 {\cal H}_{{\bf i}j\sigma} & \Delta_{\bf ij} \\
 \Delta_{\bf ij}^* & -{\cal H}_{{\bf ij}\bar{\sigma}}^*
 \end{array}\right)
 \left(\begin{array}{c}
     u_{{\bf j}\sigma}^n \\
     v_{{\bf j}\bar{\sigma}}^n
 \end{array}\right)
 = E_n
 \left(\begin{array}{c}
     u_{{\bf i}\sigma}^n \\
     v_{{\bf i}\bar{\sigma}}^n
 \end{array}\right)\;,
\end{eqnarray}
where ${\cal H}_{{\bf ij}\sigma}=-t_{\bf ij} + ( U\langle n_{{\bf
i}\sigma}\rangle - \mu ) \delta_{\bf ij}$. Here, we choose the
nearest-neighbor hopping $\langle t_{\bf ij}\rangle = t=1$ and the
next-nearest-neighbor hopping $\langle t_{\bf ij}\rangle = t'=-0.25$ to
match the curvature of the Fermi surface for most cuprate superconductors
\cite{norman63}. The exact diagonalization method to self-consistently
solve BdG equations with the periodic boundary conditions is employed to
get the $N$ positive eigenvalues $(E_n)$ with eigenvectors $(u_{{\bf
i}\uparrow}^n , v_{{\bf i}\downarrow}^n)$ and $N$ negative eigenvalues
$(\bar{E}_n)$ with eigenvectors $(-v_{{\bf i}\uparrow}^{n*} , u_{{\bf
i}\downarrow}^{n*} )$. The self-consistent conditions are
\begin{eqnarray}
\langle n_{{\bf i}\uparrow} \rangle &=&
  \sum_{n=1}^{2N}\left|{\bf u}_{\bf i}^n\right|^2 f(E_n)\;,\;
\langle n_{{\bf i}\downarrow} \rangle =
  \sum_{n=1}^{2N}\left|{\bf v}_{\bf i}^n\right|^2 [1-f(E_n)]\;,
     \nonumber \\
\Delta_{\bf ij} &=& \sum_{n=1}^{2N} \frac{V}{4} ({\bf u}_{\bf i}^n {\bf
v}_{\bf j}^{n*} + {\bf v}_{\bf i}^{n*} {\bf u}_{\bf j}^n) \tanh
(\frac{\beta E_n}{2})\;,
\end{eqnarray}
where ${\bf u}_{\bf i}^n = (-v_{{\bf i}\uparrow}^{n*}, u_{{\bf
i}\uparrow}^n )$ and ${\bf v}_{\bf i}^n = (u_{{\bf i}\downarrow}^{n*},
v_{{\bf i}\downarrow}^n )$ are the row vectors, and $f(E)=1\slash(e^{\beta
E}+1)$ is Fermi-Dirac distribution function. Since the calculation is
performed near the optimally doped regime, the filling factor,
$n_f=\sum_{{\bf i}\sigma} \langle c_{{\bf i}\sigma}^\dagger c_{{\bf
i}\sigma} \rangle /N_xN_y$, is fixed at $0.85$, i.e., the hole doping
$\delta=0.15$. Each time when the on-site repulsion $U$ or the 
temperature is varied, the chemical potential $\mu$ needs to be adjusted .

\begin{figure}[t]
\centerline{\epsfxsize=8.0cm\epsfbox{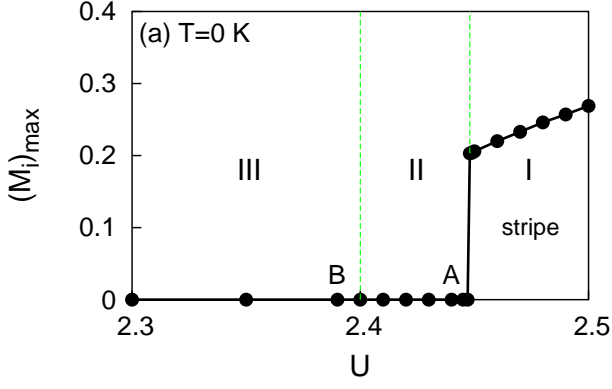}} \caption[*] {The
maximum value of the staggered magnetization $(M_{\bf i})_{max}$
as a function of $U$ at zero temperature, where $M_{\bf i}=(-1)^{\bf 
i}(n_{{\bf i}\uparrow}-n_{{\bf i}\downarrow})$. The size of the unit
cell is $N_x\times N_y = 48 \times 24$. }
\end{figure}

In Fig. 1, we plot the maximum value of the staggered
magnetization $(M_{\bf i})_{max}$ as a function of $U$ with
$V=1.0$ at $T=0$. The value of $U$ measure the strength of the
competing AF order in the background of the dSC. In region I,
where $U > U_c \sim 2.45$, the AF order with stripe modulation or
the SDW shows up and it has a period $8a$ oriented along the $x$-
or $y$-axis. The corresponding modulations in dSC and accompanying
CDW have a period $4a$. All these have been discussed in one of
our earliar works \cite{hongyi68}. In regions II and III,
where $U < U_c$, the AF order is completely suppressed by the dSC
and $(M_{\bf i})_{max}=0$, and the system is in the pure dSC
state. The transition between the stripe-modulated dSC/SDW/CDW
coexisting state and the pure dSC state is discontinuous and
of the first order. The difference between region II and region
III is that in the presence of an applied magnetic field,
dSC/SDW/CDW with stripe modulations are induced in region II while
in region III two dimensional modulations are induced around the
vortex cores. Detailed study of this issue has been given
previously \cite{hongyi0402} and will not be discussed here. In
the following, we choose $U=2.44$ (region II) which is slightly
smaller than the critical $U_c$ and $U=2.39$ (region III) to
examine their temperature effects.

\begin{figure}[t]
\centerline{\epsfxsize=8.0cm\epsfbox{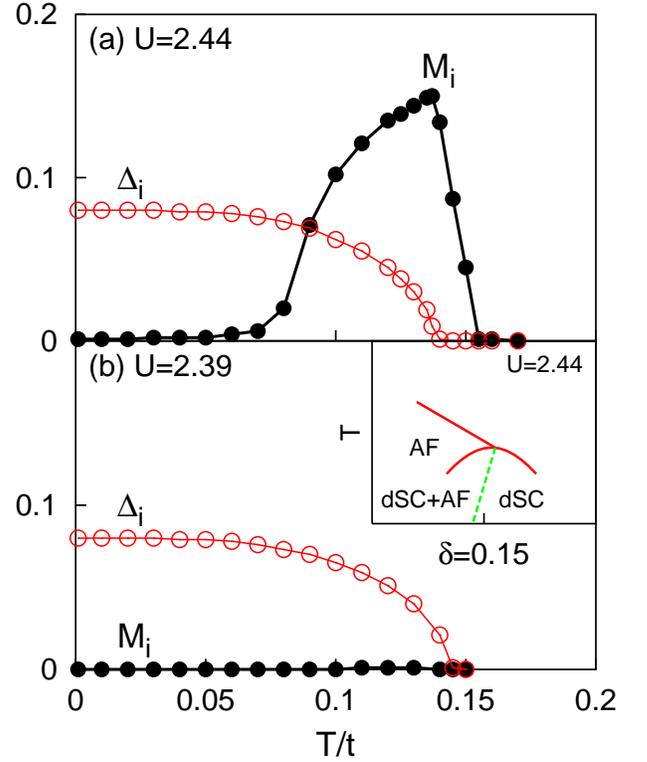}} \caption[*]{(a)
and (b) show the temperature dependence of the dSC (open circle)
and AF order (solid circle), respectively, for $U=2.44$ and
$U=2.39$. The value of the dSC order parameter $\Delta_{\bf
i}=\frac{1}{4}(\Delta_{{\bf i}+\hat{x}}+\Delta_{{\bf 
i}-\hat{x}}-\Delta_{{\bf 
i}+\hat{y}}-\Delta_{{\bf i}-\hat{y}})$ is measured in the unit of $t$ and 
$M_{\bf i}=(M_{\bf i})_{max}$.}
\end{figure}

Fig. 2 shows the temperature dependences of the dSC order
(open circle) and the staggered magnetization $(M_{\bf i})_{max}$
(solid circle) for (a) $U=2.44$ and (b) $U=2.39$. From Fig. 2(a),
it is straightforward to see that only dSC exists and the AF order
is completely suppressed at low $T$. When $T>0.06t$, the
stripe-modulated AF order and the accompanying CDW  start to
emerge, and both of them could persist above the BCS
transition temperature $T_c^{BCS}=0.14t$. This result implies that
the existence of the CDW depends critically on the presence of the
AF order, not so much on the dSC. At $T>T_c^{BCS}$, the staggered
magnetization decreases rapidly to zero at the N\'{e}el's
temperature $T_N=0.155t$. We also found that the stripe
modulation associated with the SDW has mixed periods $10a/8a$, and
those associated with the dSC and the accompanying CDW have a
mixed periods $5a/4a$ oriented either along the $x$-axis or the
$y$-axis. Within the temperature range in which the AF order appears,
these periods seem to be temperature insensitive. In Fig. 2(b),
with $U=2.39$, the AF order is completely suppressed and our system is
in the state of pure dSC at all temperatures. In both of these
cases, the dSC order parameter as a function of $T$ appears
to have the BCS-like behavior. It is interesting to note that the
$T_c^{BCS}$ in the case for $U=2.39$ [Fig. 2(b)] is slightly larger
than the one for $U=2.44$ [Fig. 2(a)]. This is because the
appearance of SDW in Fig. 2(a) at higher $T$ also suppresses the
dSC. The reason why the AF order is suppressed at low temperature
and emerges at higher temperature could be understood from the phase
diagram for $U=2.44$ near the optimal doping which is shown in the inset 
of Fig. 2(b). The feature of this phase diagram is similar to what has 
been obtained for dSC with competing $d$-density wave (dDW) order 
\cite{zhu87}.

\begin{figure}[t]
\centerline{\epsfxsize=8.0cm\epsfbox{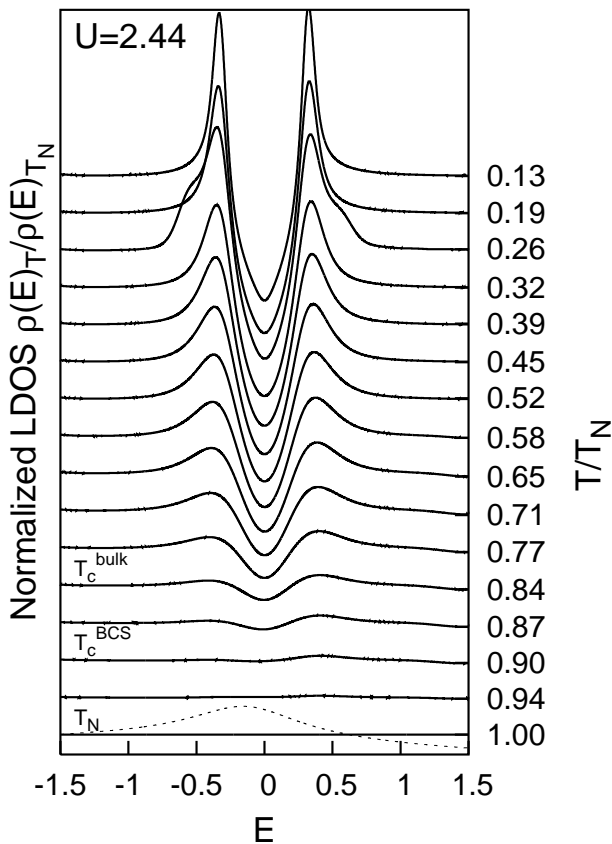}}
\caption[*]{Temperature dependence of the normalized LDOS
($\rho_{\bf i}(E)_T/\rho_{\bf i}(E)_{T_N}$, here
$T_N=0.155t$). The LDOS at $T_N$ as a function of E is
represented by the dashed line. This representative set of spectra
was shifted vertically for clarity. The wavevectors in first
Brillouin Zone are $M_x\times M_y=24\times 12$.
 }
\end{figure}

The LDOS is defined as
\begin{eqnarray}
\rho_{\bf i}(E)_T &=&-\frac{1}{M_x M_y} \sum_{n,{\bf k}}^{2N} \biggl[  
\left| {\bf u}_{\bf i}^{n,{\bf k}} \right|^2 f'(E_{n,{\bf k}}-E) \nonumber
\\  &&+\left| {\bf v}_{\bf i}^{n,{\bf k}} \right|^2 f'(E_{n,{\bf k}} + E)
\biggr]\;,
\end{eqnarray}
where $\rho_{\bf i}(E)_T$ is proportional to the local
differential tunneling conductance as measured by STM experiment,
and the summation is averaged over $M_x \times M_y$ wavevectors in
first Brillouin Zone. In order to investigate the continuous
evolution of the quasiparticle spectrum, the LDOS for $U=2.44$
from low $T$  to high  $T$ are calculated.  The
normalized LDOS which is defined as $\rho_{\bf i}(E)_T / \rho_{\bf
i}(E)_{T_N}$, are presented in Fig. 3 as
a function of energy E and at temperatures ranging from
$T/T_N=0.13$ up to $T/T_N=1.0$. It should be noticed that
$T_c^{bulk}$ ($<T_c^{BCS}$) labeled in Fig. 3 corresponds to the
temperature where the coherent peaks of the dSC flatten out, and
this signature has been used to estimate the "superconductivity
transition temperature" in many experiments. As it is indicated in
Fig. 3 that the apparent "gap" is roughly a constant up to
$T_c^{bulk}$ and progressively becomes slight larger up to $T_N$, in
consistent with the experimental measurements \cite{timusk62}.
As far as the temperature behavior of
the normalized LDOS is concerned, the essential feature of Fig. 3
displays the "pseudogap"-like characteristics. In the temperature
range of $T_c^{bulk}<T<T_N$, our system appears to be in the
"pseudogap" region according to the characterization from STM
experiments \cite{renner80,kugler86}
even though the effect due
to the phase fluctuations on the dSC order parameter
\cite{emery374,franz58} has not been taken into account. In this
region, the Fermi surface in our theory is everywhere gaped while
the ARPES experiment \cite{norman392} indicates that the gaps
occur only near ($\pm\pi,0$) and ($0,\pm\pi$). This difficulty so
far has not been solved in the existing literatures.

In order to compare with the STM experiment, we calculate the
LDOS images between the energy
$0.0$ and $0.4$ with the increment $0.01$. At the temperature
$T/T_N=0.94$, which is larger than $T_c^{BCS}$ and in the "pseudogap"
region, we found that the LDOS images have stripe modulations with
mixed periods of $5a/4a$ which is energy independent. The periods
of stripes spread into $5a-5a-4a-5a-5a$ pattern oriented along
either the $x$-axis or the $y$-axis on a $24 \times 48$ lattice. The
system seems trying to establish a periodicity that is
incommensurate with the underline lattice, but fails to do so
because the calculation is performed in a discrete and finite
lattice, not in a continuum. In Fig. 4(a), we show such an image
at $E=0.1$. The pattern in Fig. 4(a) still remains
even as the temperature drops to $T=0.08t$ [Fig.2(a)]. As it will
be shown below that an average periodicity $4.8a$ for the stripe
modulation could be assigned in this case. The $x$- and
$y$-oriented stripe modulations are degenerate in energy and it is
thus possible to use the minima of a double-well potential to
represent the states of the $x$- and $y$- oriented stripes. Since
the modulations discussed here are weak perturbations to the host
system, we do not expect that the barrier in the mid of the
double-well potential to be large and the tunneling between these
two degenerate states could be achieved easily by quantum
mechanics and thermal activation. When both $x$- and $y$-oriented
stripe-modulations show up in the time duration when the
experiment is performed, checkerboard patterns should be observed.

\begin{figure}[t]
\centerline{\epsfxsize=8.0cm\epsfbox{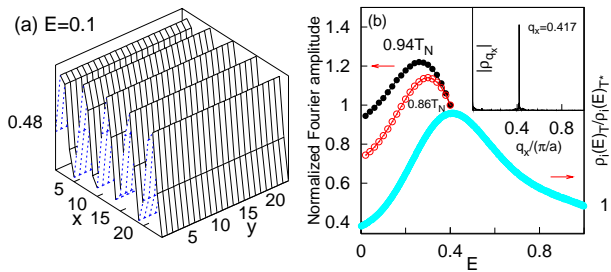}} \caption[*]{(a)
The LDOS image at $E=0.1$ and $T=0.94T_N$. (b) The energy evolutions of 
the Fourier amplitude at $T=0.94T_N$ (solid circle) and $T=0.86T_N$ (open 
circle) of the LDOS image at $q_x=0.417(\frac{\pi}{a})$, and the LDOS at 
$T=0.94T_N$. The insert in (b) is the Fourier amplitude of the LDOS image 
in (a). The wavevectors which have been used to calculate the LDOS image 
are $M_x\times M_y=12\times 12$ in first Brillouin Zone.}
\end{figure}

In order to get a further understanding of stripe modulation, let
us investigate the Fourier transform of the LDOS image
\begin{eqnarray}
\rho_{\bf q}(E)_T &=& \frac{1}{\sqrt{N_x N_y}} \sum_{\bf i} exp(i\;{\bf 
q}\cdot {\bf r_i})\cdot \rho_{\bf i}(E)_T\;.
\end{eqnarray}
The Fourier transform of the LDOS image in Fig. (4a) as a function $q_x$
is displayed in the insert of Fig. 4(b). It shows a very sharp peak around 
$q_x=0.417(\frac{\pi}{a})=\frac{2\pi}{4.8a}$, corresponding to an average 
"periodicity" $4.8a$ in real space. At $T=0.94T_N > T_c^{BCS}$ and in Fig. 
4 (b), the curve
represents the LDOS of a function of energy while the solid/open circle 
shows the energy evolution of the Fourier amplitude of the
LDOS image at $q_x=0.417(\frac{\pi}{a})$, which are normalized by their 
values at $E=0.4$. Starting from the higher energy side, the
solid circle first rises to a maximum at $E=0.27$ and then dips as
$E$ approaches to zero. This energy dependence is predicted for
the case where we only have SDW/CDW without dSC in presence. When
the temperature is lowered to $T=0.86T_N < T_c^{BCS}$ where
dSC/SDW/CDW are all in coexistence, a result represented by the
open circle is obtained for the Fourier amplitude of the LDOS image.
If one introduces an additional parameter to consider
the effect of phase fluctuations on the dSC order parameter, the
dip in the low energy side of the open circle might be removed and
flattened after reaching the maximum \cite{handong0402} and the
obtained result could agree better with the STM measurement
\cite{vershinin303}. A simplified procedure of doing this has been
previously proposed \cite{franz58} and we are not going to repeat
this again in the present paper. It also necessary to mention
here that a different work has also been done by Chen
$et\;al.$ \cite{handong0402} for rather underdoped sample (
$\delta=0.125$) to address a similar issue. Their periodicity $4a$ in the 
LDOS modulation shows up at all temperatures below the BCS transition 
temperature.

In conclusion, we have investigated the interplay between the dSC
and the competing AF/CDW orders at finite temperature for samples
close to the $\delta=0.15$ doping. With proper chosen parameters,
we show that the AF order could be completely suppressed by the
dSC at low $T$, and the LDOS image is
featureless. At higher $T$, the AF and the accompanying CDW
orders start to emerge and they persists even to temperatures
higher than ${T_c}^{BCS}$. This finding implies that the presence
of CDW depends on the existence of AF order, not so much on the
dSC. When the AF order emerges, stripe like modulations appear in
all AF/dSC/CDW orders and also in the LDOS image. It is shown that the
normalized LDOS from low $T$ to high $T$ exhibits the
"pseudogap"-like behavior. The Fourier transform of the LDOS image is
peaked around ${\bf q}=(2\pi/4.8a,0)$ indicating a periodicity
$\sim4.8a$ for $y$-oriented stripes. Moreover the energy
independent checkerboard pattern which was observed by Vershinin
$et\;al.$ \cite{vershinin303} at high $T$ but not at low $T$ could be
understood in terms of the present theory if both the $x$- and
$y$-oriented stripes contribute in the time interval when the
experiment is performed. Finally we point out that for a different
sample which may have a $U$ slightly larger than $2.44$ (see Fig.
1 in region I) or a hole-doping level slightly less than $0.15$,
stripe modulations with period $4a$ would show up at $T=0$. This
result should be consistent with the energy-independent
checkerboard patterns observed in other STM experiments
\cite{howald67}.

${\bf Acknowledgements}$: We thank S.H. Pan, J.X. Zhu and Q.
Yuan for useful comments and suggestions. This work is supported
by the Texas Center for Superconductivity and Advanced Material
at the University of Houston, and by a grant from the Robert A. Welch 
Foundation.

\end{document}